\documentclass[cp, manuscript]{copernicus}
\nolinenumbers

\begin{document}

\title{Climate and ocean circulation changes toward a modern snowball Earth}


\Author[1][tobase@jamstec.go.jp]{Takashi}{Obase} 
\Author[2]{Takanori}{Kodama}
\Author[3]{Takao}{Kawasaki}
\Author[4]{Sam}{Sherriff-Tadano}
\Author[5]{Daisuke}{Takasuka}
\Author[3]{Ayako}{Abe-Ouchi}
\Author[6,7]{Masakazu}{Fujii}

\affil[1]{Japan Agency for Marine-Earth Science and Technology, Showa-machi, Kanazawa, Yokohama, Kanagawa, 236-0001, Japan}
\affil[2]{Earth-Life Science Institute (ELSI), Institute of Future Science, Institute of Science Tokyo, I7E-315, 2-12-1, Ookayama, Meguro, Tokyo, 152-8550, Japan.}
\affil[3]{Atmosphere and Ocean Research Institute, The University of Tokyo, Kashiwa, Japan}
\affil[4]{University of the Ryukyus, Nishihara, Japan}
\affil[5]{Department of Geophysics, Graduate School of Science, Tohoku University, Sendai, Japan}
\affil[6]{National Institute of Polar Research, Tachikawa, Japan}
\affil[7]{The Graduate University for Advanced Studies, SOKENDAI, Kanagawa, Japan}

\runningtitle{Climate of Snowball}

\runningauthor{Takashi Obase}

\received{}
\pubdiscuss{} 
\revised{}
\accepted{}
\published{}


\firstpage{1}

\maketitle

\begin{abstract}
It has been hypothesized that the Earth may have experienced snowball events in the past, during which its surface became completely covered with ice. Previous studies used general circulation models to investigate the onset and climate of such snowball events. Using the MIROC4m coupled atmosphere--ocean climate model, this study examined the changes in the oceanic circulation during the onset of a modern snowball Earth and elucidated their evolution to steady states under the snowball climate. Abruptly changing the solar constant to $94 \%$ of its present-day value caused the modern Earth climate to turn into a snowball state after $ \sim1300$ years and initiated rapid increase in sea ice thickness. During onset of the snowball, extensive sea ice formation and melting of sea ice in the mid-latitudes caused substantial freshening of surface waters and salinity stratification. By contrast, such salinity stratification was absent if the duration between the change in the solar flux and the snowball onset was short. After snowball onset, the global sea ice cover and the buildup of salinity stratification caused drastic weakening in the deep ocean circulation. However, the meridional overturning circulation resumed within several hundred years after the snowball onset because the density flux by sea ice production weakens the salinity stratification. While the evolution of the oceanic circulation would depend on the continental distribution and the evolution of continental ice sheets, our results highlight the gradual growth of sea ice and associated brine rejection are essential factors for the transient evolution of the oceanic circulation in the snowball events.
\end{abstract}


\introduction 
\par
Studies have suggested that Earth has experienced major glaciation episodes, and that its surface was covered with ice during the Cryogenian Period ($720$--$635$ Ma) and the Paleoproterozoic Era ($ \sim2.4$ Ga) \citep{Hoffman2017}. The snowball Earth hypothesis is supported by multiple sources of evidence such as glacial deposits in low-latitude regions \citep{Harland1964}, banded iron formation, and cap carbonates \citep{Kirschvink1992,Kirschvink1999,Hoffman2017}. The climate forcing and the distribution of continents during past snowball Earth events differed from those of the present-day Earth. Solar luminosity is estimated to have been approximately 94\% of its present-day value during the latter stage of the Cryogenian Period based on solar models \citep{Gough1981}. Several boundary conditions for climate, including the continental distribution and the concentrations of atmospheric greenhouse gases, were changing over time and differed from the current situation. It is one crucial question why snowball Earth events occurred during the Paleoproterozoic Era and the Cryogenian Period, but not since. Climate system modeling can provide the opportunity both to understand the dynamics and interactions of climate system components, and to examine the conditions necessary for the initiation, maintenance, and termination of a snowball Earth event. 

\par
One-dimensional, latitudinal energy balance models (EBMs) of the Earth have shown that the snowball state is one equilibrium solution of the climate as a function of incoming solar flux because of the presence of ice--albedo feedback \citep{Budyko1969,Sellers1969}. Recent studies have used atmospheric general circulation models (AGCMs) to produce planetary solutions in response to insolation changes, which have different distributions of surface water with conditions that range from a snowball climate to a runaway greenhouse effect \citep{Abe2011,Rose2015,Yang2017,Kodama2021}. 
While climate models typically do not consider changes in continental ice sheets, studies involving asynchronous simulation of an ice sheet model with an EBM indicate that the ice sheet has positive feedback in the snowball solution \citep{Hyde2000,Peltier2007,Liu2010}. 

Three-dimensional ocean dynamics also impact the snowball conditions. Studies using coupled atmosphere--ocean general circulation models (AOGCMs) have shown that ocean and sea ice dynamics affect the thresholds of snowball conditions via meridional transport of sea ice and albedo feedback \citep{Voigt2010,Yang2012a,Yang2012b,Yang2012c,Liu2018}. Such studies indicated that the threshold of the incoming solar flux necessary for the onset of a snowball event in the modern Earth configuration is $89.5$--$92$ \%. Subsequent studies used AOGCMs to investigate the threshold for snowball onset under the past configuration of the Sturtian ($720$--$660$ Ma) and Marinoan ($650$--$630$ Ma) periods \citep{Poulsen2002,Voigt2011,Voigt2012,Voigt2013,Liu2013,Eberhard2023}. It was demonstrated that the necessary change in incoming solar flux for the onset of a snowball event during the Marinoan or Sturtian periods was smaller than that of the present-day configuration, indicating that initiation of a snowball event would have been easier \citep{Voigt2011,Liu2013}. On the basis of the output of a two-dimensional EBM coupled with an ice sheet model \citep{Hyde2000,Peltier2004} or an AOGCM \citep{Abbot2011,Yang2012a}, it has been suggested that a slushball is one planetary steady state that is characterized by only a small area free of sea ice in tropical regions.

The oceanic circulation during the snowball state and its changes during snowball onset and termination are important for understanding biogeochemical processes and marine sediment records. \cite{Yang2012b} conducted a modern snowball experiment including post-snowball for $\sim 200$ years, and showed the buildup of strong salinity stratification because of the extensive sea ice. \cite{Ashkenazy2014} acquired the steady-state ocean circulations in the snowball climate by utilizing a three-dimensional ocean general circulation model (OGCM) coupled with a sea--glacier model with surface atmospheric conditions from an EBM \citep{Pollard2005}. They found a meridional circulation characterized by upwelling in the tropics driven by geothermal heat flux and the gradient in sea ice thickness. AOGCM has also studied the role of ocean circulations in the deglaciation from the snowball climate. It has been shown that the sea ice and the salinity stratification of the ocean are essential factors in the timescale of snowball termination and the resumption of the meridional overturning circulation (MOC) \citep{Ramme2022,Zhao2022}.

\par
As indicated above, the threshold of incoming solar flux or atmospheric greenhouse gas concentration for snowball onset has been quantified, as has the influence of the continental distribution. However, consideration of how the MOC might change over time from snowball onset to the establishment of a steady snowball state is lacking. Comprehension of the dynamics of the oceanic circulation under the conditions of a snowball state is essential for understanding the geochemical processes and elucidating their roles both in maintaining a snowball state for millions of years and in snowball termination.
\par
In this study, we conducted steady-state experiments by reducing the solar flux under the modern continental configuration using the MIROC4m AOGCM. Our experimental design was comparable to that of previous studies on the modern snowball Earth \citep{Voigt2010,Yang2012a}. We note that the conditions for snowball onset depend on several boundary conditions such as continental distribution, solar flux forcing, and atmospheric greenhouse gas forcing. Additionally, the simulated pattern of the ocean general circulation is also affected by several boundary conditions such as the continental distribution, seabed topography, and geothermal heat flux. While these differences in the boundary conditions of our experimental design limited direct comparison with past snowball Earth events, our simulation of the modern snowball Earth elucidated the drivers of the oceanic circulation. Even so, there are challenges in simulating transient evolution of the modern sea ice to very thick sea ice in snowball climate. We address this issue by using different experimental designs in the two phases of the simulation to capture the transient evolution across snowball onset.
\par
The remainder of this paper is structured as follows. In Sect. $2$, we describe our climate model and the code changes made to the ocean model to permit thick (several hundreds of meters) sea ice in MIROC4m. We present the experimental design of this article, from the onset of the modern snowball climate and the evolution after snowball onset. Sect. $3$ present our results, as well as additional sensitivity experiments. Finally, Sect. $4$ presents discussions on implications in relation to the climate evolution of snowball Earth.

\section{Methods}
\subsection{Model}
\par
We used the MIROC4m AOGCM \citep{Hasumi2004}, which contributed to the historical and future projections of climate used in the fourth assessment report of the Intergovernmental Panel on Climate Change. The MIROC4m AOGCM consists of the CCSR-NIES AGCM (horizontal resolution is T$42$ ($2.8^{\circ}$) with $20$ vertical levels), as used in snowball planet studies \citep{Abe2011,Kodama2021}, coupled with the COCO ocean general circulation model component (horizontal resolution is $1.4^{\circ}$ $\times$ $1.0^{\circ}$ with $43$ vertical levels), as in \cite{Oka2011}. MIROC4m contributed to the Paleoclimate Modeling Intercomparison Project from the second phase to the most recent fourth phase \citep{Braconnot2006,Kageyama2021}, and it has been used in simulation of the paleoclimates of glacial periods \citep{Abe2013,Obase2019,Sherriff2021,Kuniyoshi2022} and past warm climates up to the Cretaceous with vegetation feedback \citep{oishi2021,Higuchi2021}. The equilibrium climate sensitivity of MIROC4m, defined by the global mean surface air temperature (SAT) with a doubling of the preindustrial level of atmospheric $\rm{CO}_2$, is $3.9$ K \citep{Chan2020}. One note is that the global mean surface air temperature with a doubling of the atmospheric $\rm{CO}_2$ depends on the reference states (e.g. continental distributions).
\par
A previous study identified that the albedos of snow and ice are critical model parameters regarding the threshold of snowball onset \citep{Yang2012a}. We adopted the standard albedo parameters used for present-day climate and paleoclimate simulations to ensure consistency with previous studies that used MIROC4m. The shortwave (both visible and near-infrared) albedo of sea ice $(0.5)$ and ice sheet $(0.5)$ was set at a constant value. The shortwave (visible/near-infrared) albedo of snow was defined as a function of temperature to parameterize partial snow cover at relatively high temperatures, i.e., the albedo was set at a value of $(0.85/0.65)$ for temperatures of $-5$ $^\circ$C or colder, and it was reduced linearly to a value of $(0.65/0.5)$ for temperatures up to $0$ $^\circ$C. In MIROC4m, the aging and thickness change of the snow and sea ice do not affect the albedo, which might lead to a stronger sea-ice albedo feedback in a snowball initiation. The COCO ocean model adopts a dynamic and thermodynamic sea ice model. Therefore, the maximum sea ice concentration was set as a function of sea ice thickness (linear function from $0.95$ to $1$ for ice thickness of $0$ to $12$ m) to parameterize sea ice leads. The salinity of sea ice was set to $5$ psu.

\subsection{Modification in the ocean component of MIROC4m for post-snowball simulation}
\par
The COCO ocean model uses a hybrid vertical coordinate, where the top $8$ layers ($45$ m) are represented by sigma layers to represent sea surface height changes, and the bottom layers are represented by $z$ coordinates. The formation of sea ice reduces the local sea surface height because freshwater is extracted from the ocean and incorporated in the ice during sea ice formation. This means that the sea ice thickness in a snowball climate (greater than $\sim 45$ m) leads to crashing of the ocean model because the sea surface height becomes lower than the thickness of the sigma layers. To avoid this problem, we changed the code to conserve the local sea surface associated with sea ice formation or melting. Meanwhile, the calculated ocean salinity change associated with sea ice formation or melting was considered. We incorporated this change to ensure minimal difference in the climate state of the present-day simulation. Under this setting, both global water volume (sum of seawater and sea ice) and total salinity content were no longer preserved. Nevertheless, the effect of brine rejection associated with sea ice formation and freshening associated with sea ice melting were represented as in the original model. Note that the global-mean salinity trends in the pre-industrial simulation in this revised model is 0.002 psu in 1000 years, which is of the same magnitude as the original simulation. As described in Sect. 3, we obtained comparable present-day sea ice distributions and meridional overturning circulations in the preindustrial simulations as generated in the original MIROC4m simulations.

\subsection{Experimental design 1: pre-snowball onset}
\par
All the experiments were initialized with the steady-state of the preindustrial simulations. The orbit of the Earth and the atmospheric greenhouse gas concentrations ($\rm{CO}_2$ concentration: $285$ ppm) were set to pre-industrial values, and the solar constant was set at $1366.12$ $\mathrm{W/m^2}$ \citep{Nozawa2007}. In experiments TSI091--TSI100, the solar constant was changed from $91$--$100$ \% of the preindustrial value, as in previous modern snowball studies \citep{Voigt2010,Yang2012a}. The geothermal heat flux at the seafloor was set to zero, as in the preindustrial and paleoclimate simulations.
\subsection{Experimental design 2: post-snowball onset}
\par
The climate turned into snowball state at year $430$ in the TSI091 experiment and at year $1330$ in the TSI094 experiment (Fig. \ref{timeevoa}). In the phase after snowball onset, we apply several changes in model settings to get reasonable snowball climates. First, the wind stress over the ocean was turned off after the snowball onset. This experimental design is in line with previous stuides, where turning off the air--sea momentum flux when the sea ice thickness exceeded $6$ meters \cite{Pollard2017} or in a deglaciation stage \cite{Ramme2022}. The sea ice is not fully stagnant in this setting; it can move due to ocean circulation. Note that the maximum sea ice velocity under globally sea ice covered snowball climate in our simulation is $\sim 2 \mathrm{cm/sec}$, which is one-tenth of the modern climate. The simulated horizontal flow of the sea ice is much faster than that in sea-glacier modeling studies \citep{Pollard2005,Tziperman2012}. We note that it is necessary to incorporate the stress balance of the sea-glacier model to provide a reasonable discussion of the horizontal flow of ice covers the global ocean. Second, we set all vegetation over the land grids with bare soil after snowball onset, while the present-day ice sheet over Greenland and Antarctica were kept unchanged. The experimental design setting bare soil rather than ice sheet is preventing infinite source of water via sublimation of the ice sheet, which impacts conservation of water in the climate model. Third, the minimum turbulence coefficient in the atmospheric boundary layer over the ocean was increased by a factor of 50. This change is applied solely to prevent a model crash, as we found the model crashed in a snowball climate very early without this modification. We found that the model crash can happen due to negative atmospheric water vapor content, caused by substantial atmospheric water vapor condensation over the sea ice. We found that substantial atmospheric water vapor is caused by an unrealistically strong atmospheric inversion layer over the ocean (i.e., a $2$-m air temperature difference of up to $20$ $^{\circ}$C relative to the surface temperature). The increase in minimum turbulence coefficient in the atmospheric boundary layer over the ocean prevents the occurrence of strong surface temperature, and enables steady-state simulations of the snowball climates.
\par
Using the modifications above, the snowball simulations (TSI094 and  TSI091) were continued, and the total simulation ran for $3000$ years for the TSI094 experiment (Table 1). Further We conducted additional experiment (TSI094SENS) to examine the role of snow albedo on the precipitation and atmospheric circulations in the snowball climate, by reducing the maximum snow albedo to 0.7. The TSI094SENS was branched at the end of TSI094 simulation, and 200-year integration was conducted 
\\

\begin{center}
\begin{tabular}{ccc} \hline
 Name & Solar constant & Simulation \\ \hline
 TSI100 & 100\% (1366.12 $W/m^2$)& 1000 years from the present-day climate \\
 TSI096 & 96\% & 2000 years from the present-day climate \\
 TSI094 & 94\% & 3000 years from the present-day climate \\
 TSI091 & 91\% & 1000 years from the present-day climate \\ 
 TSI094SENS & 94\% & reduced snow albedo, 200 years from the end of TSI094 \\
\hline
\label{table1}
\end{tabular}
\end{center}
\belowtable{Table 1: List of experiments in this study.} 

\section{Results}
\subsection{Onset of snowball climate}
\par
\begin{figure}[t]
\includegraphics[width=9cm,clip]
{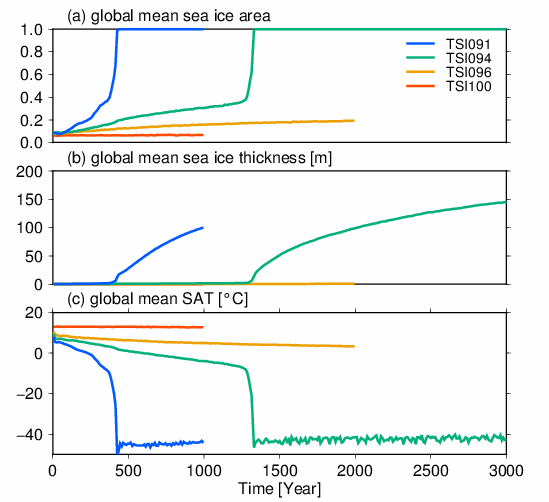}
\caption{Time series of (a) global mean sea ice covered area, (b) global mean sea ice thickness, and (c) global mean surface air temperature (SAT).}
\label{timeevoa}
\end{figure}

Fig. \ref{timeevoa} shows the time series of the global mean sea ice area, sea ice thickness, and annual SAT for all four experiments. We did not realize the snowball state in experiment TSI096 (solar constant set at $96$ \% of the present-day value) within $2000$ years (yellow line), indicating that the solar constant threshold for snowball onset was below $96$ \%. By contrast, when the solar constant was set to $94$ \% of the present-day value (TSI094), the entire ocean became covered with sea ice after approximately $1330$ years (green line). There are still gradual trends in sea ice area in the TSI096, therefore, the solar flux of $94$ \% is a sufficient condition for a modern snowball climate in MIROC4m. The solar constant required for snowball initiation in our study was larger than that reported in previous studies with modern configurations \citep{Voigt2010,Yang2012a,Yang2012b,Yang2012c,Liu2018}, suggesting greater sensitivity to insolation forcing in MIROC4m. Additionally, the sea ice area just before entering the snowball phase was approximately $40$ \%, i.e., smaller than that reported in the previous studies listed above, suggesting stronger ice--albedo feedback in MIROC4m. At the end of the TSI094 experiment, after $\sim 1700$ years from snowball onset, the global mean sea ice thickness was close to $150$ m with a trend of increase (Fig. \ref{timeevoa}b). Although the time required for snowball onset was different between the TSI091 and TSI094 experiments, the rates of sea ice thickening and fall in global mean SAT after snowball onset were the same (Fig. \ref{timeevoa}c). Moreover, the simulated sea ice growth rate of both experiments was the same magnitude to that of \cite{Ramme2022}, realizing a sea ice thickness of $60$ m in $150$ years after snowball onset.

\par
\begin{figure}[t]
\includegraphics[width=9cm]
{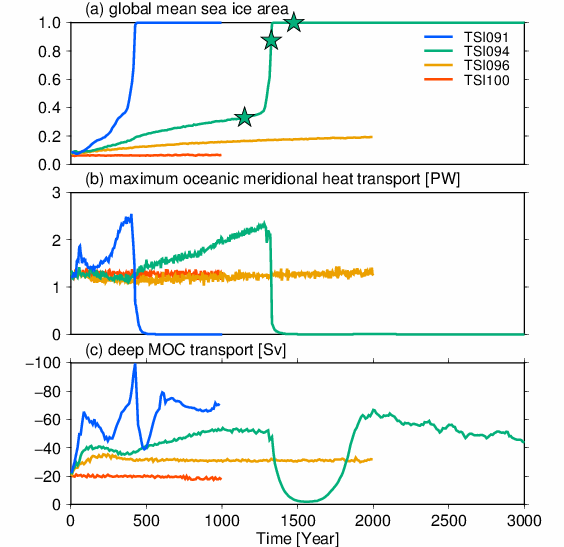}
\caption{Time series of (a) the global mean sea ice area (as in Fig. \ref{timeevoa}), (b) maximum meridional heat transport by the ocean and (c) the deep MOC streamfunction (depth of 3000 m to the seafloor) corresponding to AABW cell. Green stars in (a) represent snapshot states depicted in Fig. \ref{tsi094}.}
\label{timeevoo}
\end{figure}

Fig. \ref{timeevoo} shows time series of the meridional heat transport by the ocean and the deep MOC (associated with the Antarctic Bottom Water, AABW) cell. As in \cite{Voigt2010}, both oceanic meridional heat transport and AABW cell strengthened before the onset of snowball in the TSI091 and TSI094 experiments. However, certain differences exist between the MOC time series of the TSI091 and TSI094 experiments. The first difference is the overshoot of the AABW cell just before snowball onset. The strength of the AABW cell reached a value of approximately $100$ Sv in the TSI091 experiment, whereas no such overshoot occurred in TSI094 and the strength of the AABW cell remained at approximately $50$ Sv (Fig. \ref{timeevoo}c green). The second difference is the weakening after snowball onset. In the TSI091 experiment, the strength of the AABW cell fell to a minimum value of approximately $40$ Sv some $\sim 100$ years after snowball onset, whereupon it stabilized. In contrast, the strength of the AABW cell fell to less than $1$ Sv approximately $300$ years after snowball onset in the TSI094 case, following which it gradually recovered to reach a flow rate of $50$ Sv (similar strength to that in TSI091) some $\sim 600$ years after snowball onset.
\par
\begin{figure}[t]
\includegraphics[width=14cm]
{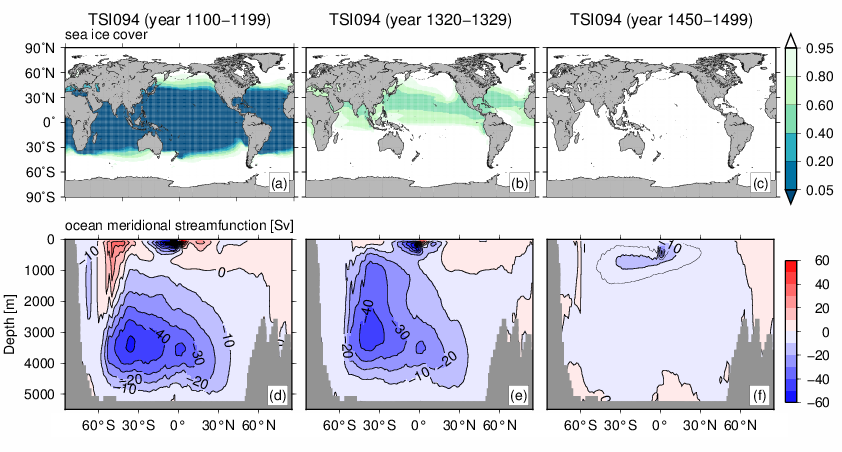}
\caption{Sea ice extent and oceanic meridional streamfunction (positive indicates clockwise circulation) at three selected time-slice snapshots from the TSI094 experiment corresponding to the green stars shown in Figure \ref{timeevoo}(a). }
\label{tsi094}
\end{figure}

Figure \ref{tsi094} shows the evolution of the sea ice extent and global MOC in the TSI094 simulation across snowball onset, from partial coverage of sea ice to full snowball conditions. The transition, characterized by the migration of the sea ice edge from the mid-latitudes to tropical regions, was very rapid (Fig. \ref{tsi094}a,b), as reported in previous studies \citep{Donnadieu2004,Marotzke2007,Voigt2010}. The streamfunction of the AABW cell was greater than that of the modern climate (Fig. \ref{tsi094}d,e) owing to the greater production of sea ice in the Southern Hemisphere than in the Northern Hemisphere. The AABW cell remained present during snowball onset. However, after snowball onset, the MOC declined rapidly and the strength of the deep ocean circulation associated with the AABW cell fell to less than $1$ Sv (Fig. \ref{timeevoo}c and Fig. \ref{tsi094}f). Even in the globally sea ice-covered state (Fig. \ref{tsi094}c), there is still atmosphere-ocean heat exchange that contributes to an increase in sea ice thickness (Fig. \ref{timeevoa}b). This gradual sea ice growth is caused by the thermal conduction within the sea ice, driven by difference between the surface and the basal temperature of the sea ice. As the sea ice thickens, the vertical temperature gradient in the sea ice becomes smaller, and the sea ice growth rate also becomes slower (Fig. \ref{timeevoa}b).

\subsection{Climate system of snowball Earth: Ocean}
\par
\begin{figure}[t]
\includegraphics[width=14cm]
{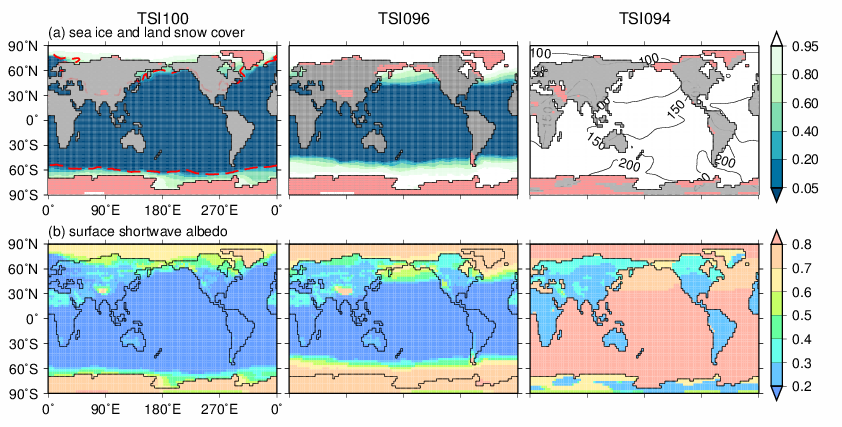}
\caption{(a) Annual mean sea ice concentration and sea ice thickness at the end of the simulations. Red lines in TSI100 indicate the present-day winter sea ice edge \citep{Hirahara2014}, and the contour in TSI094 indicates the simulated annual mean sea ice thickness (unit: m). The red shades on the land grid indicate that the monthly minimum snow depth exceeds 1 cm. (b) Annual mean surface shortwave albedo.}
\label{seaice3}
\end{figure}

We compared the sea surface climate states and ocean circulation patterns from the three experiments corresponding to the states of the modern climate (TSI100), above the snowball threshold (TSI096), and snowball Earth (TSI094). We also compared the $100$-year climatology at the end of the respective experiments (Table \ref{table1}), i.e., at year $1000$ and $2000$ in the TSI100 and TSI096 experiments, respectively. Although the sea ice was still increasing at the end of the TSI094 simulation, we analyzed the $100$-year climatology at year $3000$ from the present-day initial condition. 
The control simulation (TSI100) largely reproduced the present-day sea ice extent based on observations (Fig. \ref{seaice3}a). In TSI096, the sea ice expanded in the high latitudes of both hemispheres but the global sea ice area remained at $20$ \% (Fig. \ref{timeevoa}a). Expansion of the sea ice and prerennial snow cover over the continents in TSI096 contributed to increase in the surface albedo over the ocean (Fig. \ref{seaice3} middle). In the TSI094 experiment, the global ocean was totally covered with sea ice and snow cover, which resulted in an albedo of approximately $0.85$ (Fig. \ref{seaice3}b right). By contrast, the albedo over the continent was notably smaller , with a value of $\sim 0.3$, corresponding to that of the bare soil. This indicates an absence of perennial snow cover over most parts of the continent, as seen in (Fig. \ref{seaice3}a). Note that the surface air temperature over the continent, primarily in the Northern Hemisphere, is above $0$ $^\circ$C during boreal summer, which prevents perennial land snow cover. The sea ice thickness also exhibits a latitudinal gradient; it reaches a maximum in the Antarctic Ocean and a minimum in the Northern Pacific (Fig. \ref{seaice3}a right). 
\par
\begin{figure}[t]
\includegraphics[width=14cm]
{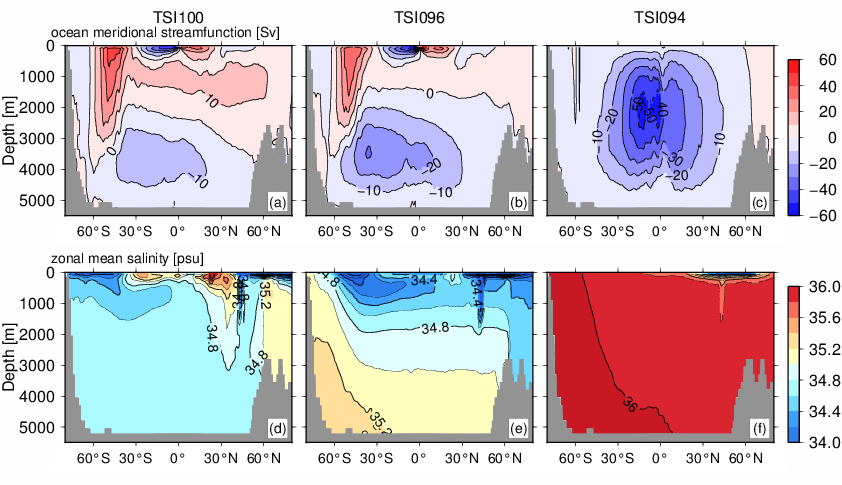}
\caption{(a) Oceanic meridional overturning circulation (positive indicates clockwise circulation) and (b) zonal mean salinity (contour interval 0.2 psu) in the three experiments.}
\label{moc3panel}
\end{figure}

The sea ice plays a critical role in simulated meridional overturning circulations. The present-day climate is characterized by cells of North Atlantic Deep Water (NADW) and AABW, as in the TSI100 simulation (Fig. \ref{moc3panel}a). The NADW cell was weakened substantially in the TSI096 case by expansion of sea ice in the North Atlantic (Fig. \ref{seaice3}a middle) which prevented deep water formation. By contrast, extensive cooling in the Antarctic Ocean promoted active sea ice production, and brine rejection led to increased AABW volume transport and more saline bottom water (Fig. \ref{moc3panel}b). In the snowball state, the meridional overturning circulation is characterized by an extensive AABW cell (Fig. \ref{moc3panel}c). 
\par
\begin{figure}[t]
\includegraphics[width=14cm]
{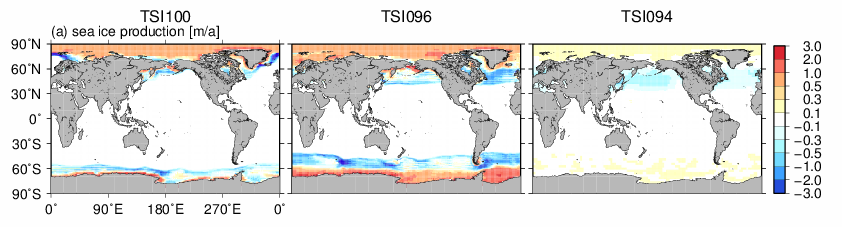}
\caption{(a) Annual mean rate of sea ice production (freshwater equivalent m/s) and (b) eastward winds at $10$-m height in the three experiments.}
\label{seaiceprod}
\end{figure}

The sea ice area was mostly in equilibrium in the TSI100 and TSI096 simulations because net sea ice production in the polar region was compensated by sea ice transport away from the polar regions and sea ice melting in subpolar regions (Fig. \ref{seaiceprod}). The TSI094 experiment exhibited slightly positive global-mean sea ice formation, as the sea ice thickened as a rate of $0.1$ m per year (Fig. \ref{timeevoa}b). The increase in sea ice volume and associated brine rejection contributed to increase in global mean salinity (Fig. \ref{moc3panel}f), as reported in previous studies \citep{Ashkenazy2014,Ramme2022}. The zonal mean salinity exhibits maximum in the Southern Ocean and along AABW cell. In contrast, the North Pacific and North Atlantic, exhibit net melting of sea ice (Fig. \ref{seaiceprod} right), contributing to smaller sea surface salinity in the Northern high latitudes (Fig. \ref{moc3panel}c, f). This meridional contrast in density flux likely reflects active deep water formation in the Southern Ocean.  
\par

\subsection{Evolution of ocean circulations}
\begin{figure}[t]
\includegraphics[width=9cm]
{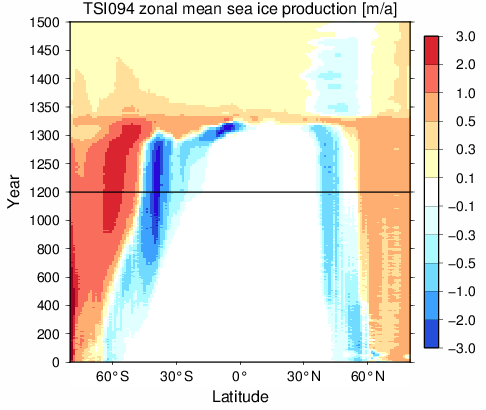}
\caption{Time series of the zonal mean sea ice production (freshwater equivalent m/a) in the TSI094 experiment. Note that the vertical scale is changed at year 1200 to highlight the rapid transition to the snowball Earth climate.}
\label{tsi094seaiceprod}
\end{figure}

Figure \ref{tsi094seaiceprod} shows the evolution of the zonal mean sea ice production in the TSI094 experiment. 
In the early phase of the experiment, sea ice formation occurred along the Antarctic coast and in the Arctic Ocean, as in the modern climate. The transport of sea ice to lower latitudes contributed to net sea ice melting in the Southern Ocean and the North Atlantic.
The sea ice edge and the area of sea ice formation migrated to lower latitudes over time as the area of sea ice increased. Notably, the Southern Ocean in the region of the Antarctic Circumpolar Current at approximately $60^{\circ}$S, which is currently an area of substantial sea ice melting, became an area of active sea ice formation after 1000 years in the experiment. The melting of sea ice in the mid-latitudes also increased, contributing to reduction in the surface salinity. Generally, greater production of sea ice in the Southern Hemisphere than in the Northern Hemisphere contributed to the strong AABW cell prior to snowball onset (Fig. \ref{tsi094}d,e) It is also notable that the rapid expansion of sea ice before snowball onset was faster in the Southern Hemisphere than in the Northern Hemisphere.

\begin{figure}[t]
\includegraphics[width=10cm]
{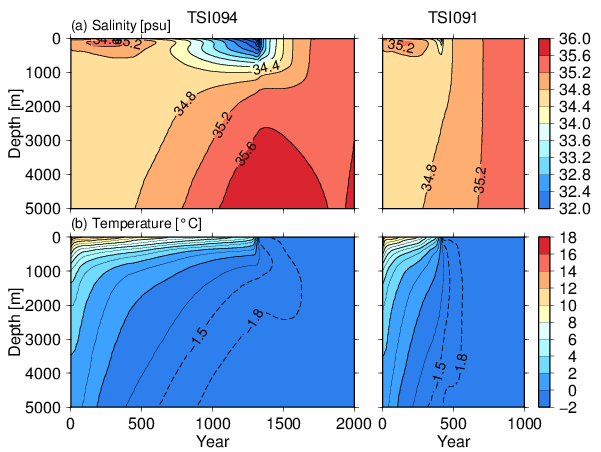}
\caption{Time series of the vertical profile of the global mean ocean (a) salinity and (b) temperature in the TSI094 and TSI091 experiments.}
\label{tsi094toso}
\end{figure}

\begin{figure}[t]
\includegraphics[width=9cm]
{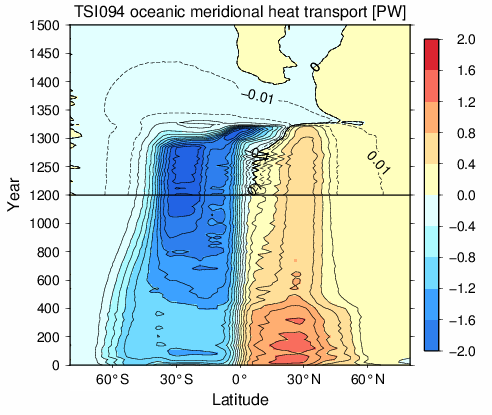}
\caption{Time series of the oceanic meridional northward heat transport in the TSI094 experiment. Note that the vertical scale is changed at year 1200 to highlight the rapid transition to the snowball Earth climate.}
\label{tsi094ohttp}
\end{figure}

Figure \ref{tsi094toso} compares Hovm\"{o}ller diagrams of the global mean ocean temperature and the salinity profile of the TSI094 and TSI091 experiments. In the TSI094 experiment, the salinity of the surface layer decreased owing to increased melting of sea ice in mid- and low-latitude areas, while the salinity of the deep layer increased owing to sea ice production and brine rejection in the polar regions (Fig. \ref{tsi094seaiceprod}). The global mean sea surface salinity dropped to $31.4$ psu in TSI094 before snowball onset (Fig. \ref{tsi094toso}a left). Conversely, the enhanced salinity stratification found in the TSI094 experiment was absent in the TSI091 experiment, where the global mean sea surface salinity at snowball onset was $33.8$ psu. The temperature and salinity profiles approached uniform values after the snowball onset in both experiments, but TSI094 took longer to remove the salinity stratification at snowball onset. Hence, the strong salinity stratification resulting from active sea ice melting just prior to snowball onset was found only in TSI094. The strong salinity stratification at snowball onset probably caused the strength of the AABW cell to weaken to approximately $1$ Sv after snowball onset in TSI094 (Fig. \ref{timeevoo}c), while the strength of the AABW cell was maintained at a strong value in TSI091 where such strong salinity stratification was absent. 
The poleward oceanic heat transport became stronger toward snowball onset (Fig. \ref{tsi094ohttp}). The increase in the meridional heat transport that occurred only in the Southern Hemisphere, similar to the results of \cite{Voigt2010}, was due to strengthening of the AABW cell and the disappearance of the NADW. After snowball onset, the oceanic heat transport diminished rapidly as the ocean temperature approached uniformity at the freezing point, eliminating the temperature difference between low- and high-latitude areas (Fig. \ref{tsi094toso}b).

\subsection{Atmosphere of the snowball Earth climate}
\par
\begin{figure}[t]
\includegraphics[width=12cm]
{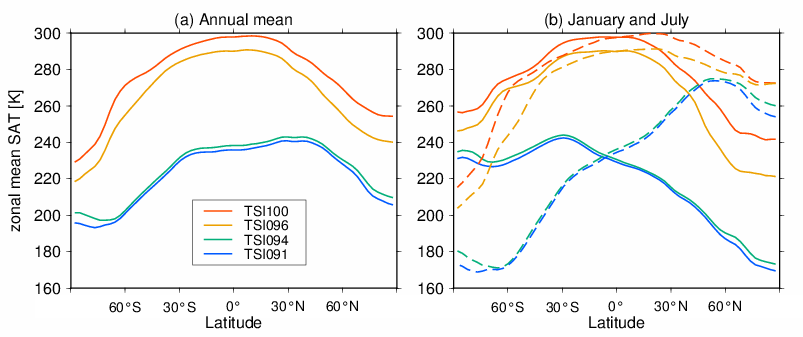}
\caption{Zonal mean $2$-m air temperature in the four experiments: (a) annual mean and (b) January mean (bold lines) and July mean (dashed lines)}
\label{zonmeant2}
\end{figure}
\begin{figure}[t]
\includegraphics[width=12cm]
{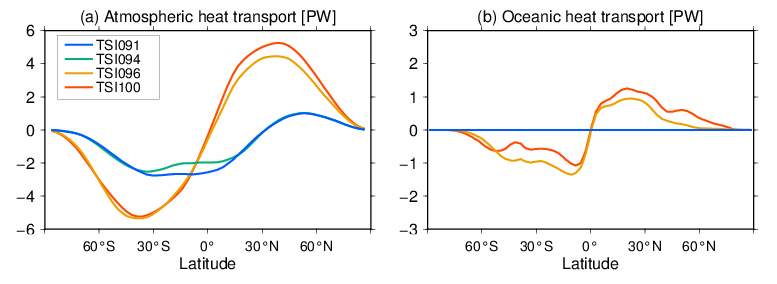}
\caption{(a) Atmospheric and (b) oceanic meridional heat transport in the four experiments.}
\label{http4}
\end{figure}

We also analyzed the atmospheric temperature and circulations at the same period (Figs. \ref{seaice3}--\ref{seaiceprod}) to elucidate the importance of air--sea fluxes as drivers of ocean circulations. The annual mean zonal mean air temperature of the snowball climate was $240$ K in the low-latitudes and $200$ K in polar regions (Fig. \ref{zonmeant2}a). The maximum monthly air temperature occurred in the summer over the Northern Hemisphere, where simulated summer air temperature over the continents were above $0 ^\circ $C (Fig. \ref{zonmeant2}b), attributed to the smaller surface albedo over the continents (Fig. \ref{seaice3}b). The smaller temperature gradient between the tropics and the polar regions caused smaller meridional atmospheric and oceanic heat transport. The oceanic meridional heat transport was weakened in TSI096 (Fig. \ref{http4}b), likely related to the absence of NADW formation (Fig. \ref{moc3panel}b), whereas in the snowball climate (TSI094 and TSI091), there was minimal oceanic heat transport (Fig. \ref{http4}). The atmosphere still transported heat but the amount was reduced markedly relative to that of the modern climate. The atmospheric circulation in a snowball climate has been investigated by multi-model study \cite{Abbot2012,Abbot2013}, with constant surface albedo $0.6$ and two atmospheric $\rm{CO}_2$ concentrations. Our snowball Earth results exhibited generally colder surface air temperatures than multi-model study, which can be explained by the different concentration of atmospheric $\rm{CO}_2$ used, and the higher albedo of snow ($0.85$) in the snowball states (Fig. \ref{seaice3}b right). 
\par
\begin{figure}[t]
\includegraphics[width=12cm]
{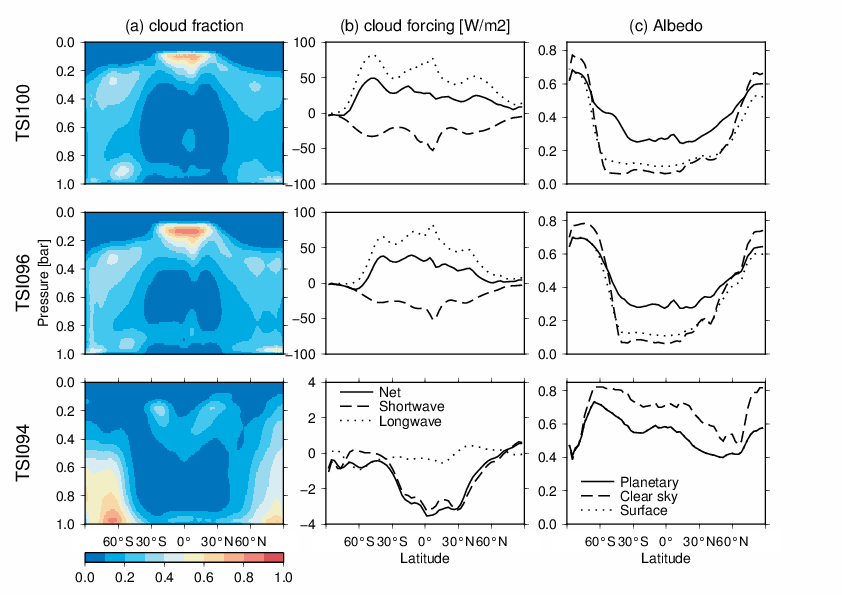}
\caption{Zonal mean cloud fraction, cloud radiation effect for shortwave and longwave radiation, and zonal mean planetary albedo, clear sky albedo, and surface albedo in the TSI100, TSI096, and TSI094 experiments.}
\label{rev13}
\end{figure}
\begin{figure}[t]
\includegraphics[width=12cm]
{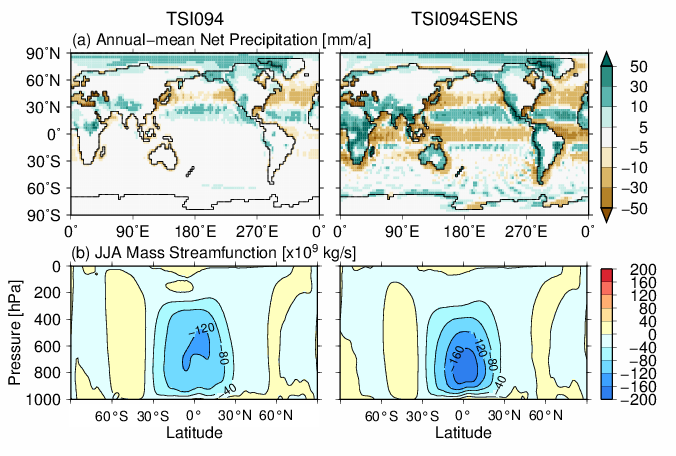}
\caption{(a) Net precipitation, defined by precipitation minus evaporation (includes ice sublimation), (b) boreal summer meridional atmospheric mass streamfunction ($10^9kg/s$, positive indicates clockwise circulation) in the TSI094 and TSI094SENS experiments.}
\label{rev14}
\end{figure}

The cloud cover in our simulation exhibited cloud fraction in the troposphere of high-latitudes (Fig. \ref{rev13}a bottom), in accordance with some models in a multi-model study with smaller atmospheric $\rm{CO}_2$ concentrations \citep{Abbot2012}. Our simulation exhibits mostly weaker cloud radiative forcing of the snowball climate among multi-model (Figure \ref{rev13}b bottom), but the magnitude of cloud radiative forcing is within ranges of the six models \citep{Abbot2012}. The net precipitation exhibits sublimation of sea ice near the equator and precipitation in mid- and low-latitudes (Fig. \ref{rev14}a left). Substantial positive net precipitation occurs in mountainous areas across the continent, whereas inland areas far from the ocean experience very sparse precipitation. 

While multimodel study showed net sublimation of sea ice near equator and stronger Hadley Circulations than the modern climate \citep{Abbot2013}, our TSI094 showed net sublimation near equator is very weak and the strength of the Hadley Circulation is weaker than the modern simulation (Fig. \ref{rev14} left). 
In the additional experiment with reduced snow albedo (TSI094SENS), more significant sublimation is exhibited in the equatorial region and net precipitation in the low latitudes (Fig. \ref{rev14}a). The Hadley circulation was also strengthened (Fig. \ref{rev14}b). Overall, the simulated precipitation pattern and Hadley circulation were closer to the multimodel study \citep{Abbot2013}, suggesting that spatial heterogeneity in the surface albedo can lead to atmospheric circulation that differs from that obtained in multi-model studies using a constant surface albedo \citep{Abbot2013}.

\section{Discussion}
\subsection{Threshold of snowball onset}
\par
On the basis of our MIROC4m experiments, we found that $94$ \% of the present-day value of insolation is one sufficient condition for the modern Earth. Previous AOGCM studies show larger reduction in solar flux in sufficient modern snowball conditions, e.g., $89.5$--$92$ \% \citep{Voigt2010,Yang2012a,Yang2012b,Yang2012c}). Additionally, the global mean sea ice area before snowball onset was $\sim 40$ \%, which is also smaller than that used in the other AOGCM studies listed above (i.e., $55$--$76$ \%). One possible explanation for the greater susceptibility of MIROC4m model to the snowball climate in response to insolation change is the albedo of snow. In MIROC4m, the albedo of snow can be as high as $0.85$ in a colder climate (Fig. \ref{seaice3}b), which is near the maximum value adopted in the AOGCMs used in previous studies \citep{Yang2012a}. One note is that the maximum snow albedo of $0.85$ is the same as recent ICON-ESM \citep{Ramme2022}.

\subsection{Evolution of ocean circulation across snowball onset}
\par
At the end of the simulation up to $1600$ years from the onset of the snowball state, active ocean circulations of approximately $40$ Sv were maintained. Although the volume transport of the MOC is of the same order of magnitude, the circulation pattern is different from that of previous study using paleotopography \citep{Ashkenazy2014}, which shows a symmetric circulation about the equator using paleotopography.
As shown by \cite{Yang2012b}, the TSI094 experiment showed that a strong salinity stratification develops just before the snowball onset. Furthermore, it was shown that this strong salinity stratification substantially weakens the deep-ocean circulation after the snowball onset. 
We also found that the strong salinity stratification at snowball onset is present in the TSI094 experiment but absent in the TSI091 experiment, suggesting that the presence of salinity stratification before snowball onset depends on the rate of cooling. If the rate of cooling was fast and the climate turned into the snowball state in a relatively short time (TSI091), the strong stratification at snowball onset was absent. By contrast, if the transition to the snowball state was gradual, the strong salinity stratification could develop before snowball onset. This would explain why the salinity stratification was not observed in \cite{Ramme2022}, who found that it took approximately $150$ years for snowball onset. It required several hundred years to resolve the salinity stratification in our TSI094 experiment (Fig. \ref{tsi094toso}a left) as opposed to TSI091 and the experiments of \cite{Ramme2022}. 
After the snowball onset, the MOC experienced gradual recovery within several hundred years. The MOC changes across the snowball transition in TSI094 can be explained as follows based on our analysis. At snowball onset, the strongly stratified ocean, mainly by salinity leads to an initial, abrupt weakening of the MOC. After the snowball onset, the gradual sea ice production and associated brine rejection erode the strong ocean stratification (Fig. \ref{tsi094toso}a). At the same time, atmospheric conditions in the North Pacific region remain relatively warm, favoring sea ice melting (Fig. \ref{seaiceprod}). In contrast, the Southern Ocean remains relatively cold, promoting active sea ice formation, generating a meridional density gradient (Fig. \ref{moc3panel}f) that ultimately drives the gradual resumption of the AABW cell (Fig. \ref{moc3panel}c). These results suggest that atmospheric conditions at the sea surface, shaped by continental configuration and the extent of land ice sheets, can strongly influence the temporal evolution and the steady-state of the MOC in a snowball climate.

\subsection{Potential limitation on ocean circulation}
\par
In MIROC4m, the surface momentum flux between the atmosphere and the ocean under the presence of sea ice is formulated using a nondimensional drag coefficient which is independent of sea ice thickness. We abruptly turned off the air-sea momentum flux after the snowball onset to get reasonable ocean circulation in the snowball climate, but the actual change would be gradual. While this experimental design is still in line with previous study addressing across snowball onset, \citep{Pollard2017,Ramme2022}, there is still knowledge gap of how to formulate or parameterize the stresses and dynamics of thick sea ice over the ocean for fully transient simulation across snowball onset. Improvements in the model would include extension of the Antarctic ice sheet and ice shelf model to represent the dynamics of globally covered sea ice. Other issues are related to the boundary conditions used in the climate model. Based on the balance between the vertical diffusion of heat in the sea ice and the typical value in the geothermal heat flux of the Earth, the sea ice thickness would reach a steady state of approximately $1000$ m where the vertical diffusion of heat in the sea ice and the amount of geothermal heat flux are balanced. As shown by \cite{Ashkenazy2014}, the upwelling ocean circulation at the equator is achieved via the inflow of geothermal energy and the thickness of the sea ice becoming thinner toward the equator. The inclusion of geothermal heat flux and bathymetry may lead to different MOC.

\subsection{Potential limitation of atmosphere}
\par
In case of reduced snow albedo (TSI094SENS), the value of net precipitation was approximately $10$ mm per year in the mid-latitude (Fig. \ref{rev14}a right), which is in accord with the result derived from the multimodel AGCMs used for snowball simulations \citep{Abbot2013}. The results of additional experiments suggested that the spatial distibution of surface albedo could have considerable impact on the climate of snowball Earth via the precipitation, and potentially mass balance of the continental ice sheets. Furthermore, not all land surfaces were covered with perennial snow cover in snowball climate in our TSI094 simulation (Fig. \ref{seaice3}a). This is consistent with a study using the same AGCM as this study \citep{Abe2011}, which found that land snow cover requires a lower solar flux than that required for global sea ice cover. Substantial land-snow free area of snow cover can be found in previous studies with the modern snowball \citep{Yang2012a} and paleotopography \citep{Benn2015}, though the atmospheric $\rm{CO}_2$ concentration were not necessary the same. The lower albedo of the land due to the absence of snow cover contributes to the warmer surface temperature and the sublimation of ice, which contributes to positive feedback for further warming over the land. We note that MIROC4m tends to exhibit warmer summer surface air temperature over the land, therefore the extent of land-snow free area would depend on climate models used. 
\subsection{Other climate system feedbacks and implications}
\par
In the MIROC4m simulations, the extent of the continental ice sheet and vegetation was assumed the same as that of the present-day in all simulations, meaning that the continental ice sheet was limited in Greenland and Antarctica. Although a standard AOGCM would not forecast changes in the ice sheet because such changes take much longer than the typical period of an AOGCM simulation, it has been shown that the prescription of the ice sheet can change the threshold for snowball onset \citep{Liu2017}. Ice sheet modeling studies with one-way, offline ice sheet model simulations using AGCM climate fields \citep{Donnadieu2003,Pollard2004} or ice sheet models asynchronously coupled with an AGCM \citep{Benn2015} found that the development of an ice sheet in low-latitude areas is possible because of net precipitation. Additionally, results derived from an ice sheet model coupled with an EBM indicated that the continental distribution affects the solution of the snowball state because of migration of ice sheet margins \citep{Liu2010}. The extent and evolution of ice sheet with coupled climate--ice sheet model in future studies are necessary to examine the geological records \citep{Donnadieu2003,Pollard2004,Benn2015}, and these climate states are crucial in quantifying the threshold of deglaciation from the snowball state \citep{Pierrehumbert2004,Pierrehumbert2005,LeHir2007,Hu2011,Abbot2012,Abbot2013,Abbot2014,Wu2021,Zhao2022}. 

\par
We note that several boundary conditions were different from the actual situation of snowball events during the Cryogenian Period. First, we used the modern configuration of the continental distribution. Previous studies showed that the reduction in solar flux necessary for snowball onset was smaller than that associated with the present-day continental configuration, indicating that initiation of a snowball climate is easier if paleotopography is considered \citep{Voigt2011,Liu2013}. Second, we conducted simulations with incoming solar flux, but changes in atmospheric greenhouse gas forcing would produce different results because the distribution of radiative forcing would be different. Specifically, \cite{Liu2013} estimated the threshold of atmospheric $\rm{CO}_2$ as $80$ ppm for the Sturtian and $150$ ppm for the Marinoan configuration, and it can change by $\sim30$ ppm depending on the optical depth of the aerosol parameterisation. \cite{Feulner2014} estimated the threshold of atmospheric $\rm{CO}_2$ concentration as $100-110$ ppm for Sturtian and $120-130$ ppm for Marinoan configurations, respectively. While this study clarified the timescale of the evolution of deep ocean circulation across the snowball onset, the verification of the three-dimensional ocean ciruclation is difficult because there was no snowball events in the modern configuration. A future study utilizes Sturtian and Marinoan configurations to understand the evolution of the deep-ocean circulation. Moreover, the inclusion of geochemical processes in climate models can be verified by geological and geochemical records \citep{Hoffman2017}.
\par
Our simulations prescribe atmospheric greenhouse gas concentrations as input parameters. In contrast, three-dimensional OGCM studies showed that global cooling above the threshold of snowball onset contributes to oceanic carbon uptake and reduces the concentration of atmospheric $\rm{CO}_2$ by $50$--$150$ ppm \citep{Oka2011,Liu2023}, suggesting that the oceanic carbon cycle in response to cooling acts as positive feedback on snowball onset. 
The TSI094 results indicated that the ocean can have very strong salinity stratification before snowball onset if the conditions of the solar flux are just below the threshold (Fig. \ref{tsi094toso}a). In our simulation, the concentration of atmospheric $\rm{CO}_2$ was set at a constant value during the $1000$ years before snowball onset when the strong salinity stratification developed. If the model forecasts the marine carbon cycle and the concentration of atmospheric $\rm{CO}_2$, the strong salinity stratification would uptake carbon in the deep and bottom ocean, as was found in both \cite{Oka2011} and \cite{Liu2023}, thereby lowering the concentration of atmospheric $\rm{CO}_2$ and possibly causing snowball onset before the strong salinity stratification developed. It is also noted that the current continental distribution enables active sea ice production in the region of the Antarctic Circumpolar Current (Fig. \ref{tsi094seaiceprod}), which might contribute to the formation of strong salinity stratification, and this should be examined using the past continental distribution. Additionally, the presence of ice sheets and the weathering rates associated with the past continental distribution could also cause changes in the concentration of atmospheric $\rm{CO}_2$ \citep{Tajika2003,Donnadieu2004b,Peltier2007,LeHir2009}. 
\par
Despite the limitations in the models and experimental design adopted in this study, our simulations provide insights into the role of the oceanic circulation in the climate state before snowball onset, during snowball onset, and during a snowball Earth event. If the reduced solar flux was slightly weaker than the threshold required for snowball onset, active sea ice production in the polar regions and associated brine rejection would increase the salinity of the deep ocean, while melting of sea ice in the mid--low latitudes would reduce the salinity of the surface ocean, resulting in strong stratification. This suggests that in intensive glacial periods shorter than a snowball event, there would be strong ocean salinity stratification because of extensive sea ice cover.
Once the forcing or concentration of atmospheric $\rm{CO}_2$ associated with ocean carbon uptake exceeded the threshold for snowball onset, the entire globe would become covered in sea ice within a few decades, as reported in previous studies. After snowball onset, even with the current model settings of substantial air--sea flux, it would take several hundred years to eliminate the strong ocean stratification and to achieve a steady MOC with nearly uniform temperature and salinity fields. Our findings highlight the necessity for evaluating the impact of ocean circulation changes on geochemical modeling, including continental ice sheet evolutions.




\codedataavailability{The MIROC4m code associated with this study is available to those who conduct collaborative research with the model users under license from copyright holders. All model data supporting our findings will be archived at Zenodo.} 










\authorcontribution{TO and TK conceived the study. TO and TK improved the model for use in this study. TO, TK, and SS-T analyzed the data. TO wrote the manuscript with input from all coauthors.} 

\competinginterests{The authors declare that they have no competing interests.}  


\begin{acknowledgements}
We appreciate Dr. Yonggang Liu and one anonymous reviewer for their valuable comments, which have substantially improved our paper. We also appreciate the fruitful discussions conducted with Joseph Kirschvink, Ryouta O’ishi, and Tatsuo Suzuki. This research is supported by JSPS Kakenhi 23K17709, 24K17122, and a “Strategic Research Projects” grant (No. 2024-SRP-03 (TK)) from the Research Organization of Information and Systems. TO receives additional support from JPMXD0722681344. This study is also supported by grants from the Astrobiology Center of the National Institutes of Natural Sciences (No. AB0608) and the Itoh Science Foundation (TK). The simulations were performed on the Earth Simulator 4 at the Japan Agency for Marine-Earth Science and Technology (JAMSTEC). We thank James Buxton MSc, from Liwen Bianji (Edanz) (www.liwenbianji.cn/), for editing the English text of a draft of this manuscript. 
\end{acknowledgements}



\bibliographystyle{copernicus}
\bibliography{reference.bib}

\end{document}